\documentclass[tightenlines,aps,showpacs,date,nofootinbib]{revtex4}

\include{graphics}

\usepackage{float}
\usepackage{amsmath,amsfonts,graphicx,bm}

\newcommand{\ba}{\begin{eqnarray}}
\newcommand{\ea}{\end{eqnarray}}
\newcommand{\be} {\begin{equation}}
\newcommand{\ee} {\end{equation}}
\newcommand{\GeV}{\mbox{\rm GeV}}
\newcommand{\MeV}{\mbox{\rm MeV}}
\newcommand{\order}{{\cal O}}

\newcommand{\psibar}{\overline{\Psi}}

\begin{document}

\title{Unquenched determination of the kaon parameter 
$B_K$  from improved staggered fermions}

\author{Elvira G\'amiz, Sara Collins, Christine T.H.~Davies}
\affiliation{Department of Physics \&
              Astronomy, University of Glasgow, Glasgow, G12 8QQ, UK}
\email{e.gamiz@physics.gla.ac.uk,c.davies@physics.gla.ac.uk}
\author{G. Peter Lepage}
\affiliation{Laboratory of Elementary-Particle Physics,
Cornell University, Ithaca, New York 14853, USA}
\author{Junko Shigemitsu}
\affiliation{Physics Department, The Ohio State University, Columbus, 
OH 43210, USA}
\author{Matthew Wingate}
\affiliation{Institute for Nuclear Theory, University of Washington, Seattle, 
WA 98115, USA}

\author{HPQCD and UKQCD collaborations}

\begin{abstract}
The use of improved staggered actions (HYP, Asqtad) 
has been proved to reduce the scaling corrections that affected 
previous calculations of $B_K$ with unimproved (standard) staggered 
fermions in the quenched approximation. This improved 
behaviour allows us to perform a reliable calculation
of $B_K$ including quark vacuum polarization effects, using the
MILC configurations with $n_f=2+1$ flavours of sea fermions. 
We perform such a calculation for a single lattice spacing, $a=0.125~fm$,  
and with kaons made up of degenerate quarks with $m_s/2$.  
The valence strange quark mass $m_s$ is fixed to its physical value and 
we use two different values of the light sea quark masses. After a chiral 
extrapolation of the results to the physical value of the sea quark masses,  
we find $\hat B_K = 0.83\pm0.18$, where the error is dominated by the 
uncertainty in the lattice to continuum matching at $\order (\alpha_s^2)$. 
The matching will need to be improved to get the precision 
needed to make full use of the experimental data on $\varepsilon_K$ 
to constrain the unitarity triangle.
\end{abstract}

\maketitle


\section{Introduction}

In the last few years lattice calculations have started achieving the level of 
precision and the control of uncertainties necessary to extract 
phenomenologically relevant results \cite{Davies:2003ik}. Simulations that 
incorporate quark vacuum polarization effects 
are required for this task, since the uncontrolled errors 
associated with the quenched approximation are usually the main source of 
uncertainty in these calculations. An example of this can be seen in the 
study of indirect CP violation in the neutral kaon system.

The CP violating effects in $K^0-\overline{K^0}$ mixing are parametrized by 
$\varepsilon_K$, which is defined in terms of the decay amplitudes 
of the physical neutral kaon states $K_L$ and $K_S$ into a state of two 
pions with isospin equal to 0 as
\be
\varepsilon_K = \frac{A(K_L\rightarrow (\pi\pi)_{I=0})}
{A(K_S\rightarrow (\pi\pi)_{I=0})}\,.
\ee
Experimentally, this quantity is known with a precision of a 
few percent. On the other hand, the theoretical calculation in the 
Standard Model, under reliable assumptions, yields the expression
\be
\varepsilon_K \simeq \frac{e^{i\pi/4}} {\sqrt{2}\,\Delta M_K}
{\rm Im} M_{12}\, ,
\ee
where $\Delta M_K$ is the mass difference between $K_L$ and $K_S$, a very 
well measured quantity, and $M_{12}$ is defined by
\be
2 m_K M_{12}^*=\langle \overline{K^0}|H_{eff}^{\Delta S=2}|K^0\rangle\,.
\ee
$\varepsilon_K$ is thus determined by the hadronic matrix 
element between $K^0$ and $\overline{K^0}$ of the $\Delta S=2$ 
effective Hamiltonian
\be\label{effHam}
H_{eff}^{\Delta S=2}= C_{\Delta S=2}C(\mu)\int d^4x \,Q_{\Delta S=2}(x)
\ee
with
\be
Q_{\Delta S=2}(x) = \left[\bar s_{a}\gamma_{\mu}
d_{a}\right]_{V-A}(x)
\left[\bar s_{b}\gamma^{\mu}d_{b}\right]_{V-A}(x).
\ee

The Wilson coefficient $C(\mu)$ is a perturbative 
quantity known to NLO in $\alpha_s$ in both the Naive Dimensional 
Regularization (NDR) and the 't Hooft-Veltman (HV) schemes. 
The coefficient $C_{\Delta S=2}$ includes known functions of the masses 
of particles that have been integrated out, and also depends on a 
certain combination of Cabibbo-Kobayashi-Maskawa (CKM) matrix elements.  
It is these latter elements about which we would like to obtain information. 
The combination of CKM parameters entering in $C_{\Delta S=2}$ is  
${\rm Im}\left(V_{ts}V_{td}^*\right)^2$, 
which is equivalent to the unitarity triangle (UT) combination of parameters 
$\overline{\eta}\left[\left(1-\overline{\rho}\right)+{\rm const.}\right]\,$ 
\cite{UTdefinitions}.
For reviews where explicit expressions for the coefficients in 
(\ref{effHam}) can be found, see \cite{Buras}. 
The two-loop expression for the Wilson coefficient $C(\mu)$ is given by 
\be\label{Wilsoncoeff}
C(\mu) = \left(1+\frac{\alpha_{\overline{MS}}(\mu)}{4\pi}
4\left[\frac{\gamma_1}
{\beta_0}-\frac{\beta_1\gamma_0}{\beta_0^2}\right]\right)
\left[\alpha_{\overline{MS}}(\mu)\right]^{\gamma_0/\beta_0}\, ,
\ee
where $\gamma_0$, $\gamma_1$ are the $\Delta S=2$ anomalous dimension at 
one-loop and two-loop respectively, and $\beta_0$, $\beta_1$ are the 
first two coefficients of the QCD beta function. $\gamma_0=1$ is 
independent of the number of flavours $n_f$, while the other quantities 
depend on $n_f$ in the following way
\be\label{betas}
\beta_0=-\frac{1}{2}\left(11-\frac{2n_f}{3}\right)\, ;\quad\quad
\beta_1=-\frac{1}{8}\left(102-\frac{38n_f}{3}\right)\, ;\quad\quad
\gamma_1^{\overline{MS}-NDR} = \frac{1}{16}\left(-7+\frac{4n_f}{9}\right)\,;
\ee
where only the NLO coefficient $\gamma_1$ depends on the choice of scheme and 
it is given in the $\overline{MS}(NDR)$ scheme we are going to use to quote 
the results in this work. The other coefficients are universal. 
Although $C(\mu)$ depends analytically on the number of flavours through 
the parameters in (\ref{betas}), the numerical dependency on $n_f$ 
in the conversion from the $\overline{MS}(NDR)$ scheme value 
$B_K^{\overline{MS}-NDR}(2\GeV)$ to $\hat B_K$ is negligible 
\cite{Becirevic02}. 

The matrix element 
$\langle \overline{K^0}\vert Q_{\Delta S=2}\vert K^0\rangle$, that encodes the 
non-perturbative physics of the problem, is usually normalized by 
its Vacuum Insertion Approximation (VIA) value, defining $B_K$ as the ratio
\be\label{BKdef}
B_K(\mu) \equiv \frac{\langle\overline{K^0}|
Q_{\Delta S=2}(\mu)|K^0\rangle}
{\frac{8}{3}\langle \overline{K^0}|\bar s\gamma_{0}\gamma_5d|0\rangle
\langle 0|\bar s\gamma_{0}\gamma_5d|K^0\rangle}.
\ee

The renormalization group invariant form of $B_K$, the so-called 
$\hat B_K$, is often used to quote results. It is defined 
as
\be\label{hatBk}
\hat B_K = C(\mu)B_K(\mu)\, , 
\ee
with $C(\mu)$ the Wilson coefficient in the effective 
Hamiltonian (\ref{effHam}), given at two-loops in (\ref{Wilsoncoeff}).

One can try to constrain the value of the combination of CKM matrix 
elements in $C_{\Delta S=2}$ using the experimental value of $\varepsilon_K$ 
and a theoretical calculation of $B_K$. That constraint gives a  
hyperbole in the $\rho-\eta$ plane, where $\rho$ and $\eta$ are the usual 
UT parameters \cite{UTdefinitions}. The largest source of 
uncertainty in the final results comes from  
the error in the determination of $B_K$ \cite{Charles:2004jd,Nierste05}.  
Improvement in the calculation of this kaon parameter, to 
reach at least the same level of precision (a few percent) as the rest of 
the errors entering in the analysis of $\varepsilon_K$,   
is thus crucial in order to get information about the UT.
In addition, $\varepsilon_K$ can be a very powerful probe of new physics 
\cite{Nir05}.

Due to the phenomenological relevance of $B_K$ its calculation has been 
addressed many times using different techniques. The three main continuum 
QCD-based approaches that have been used in its determination 
are QCD-Hadron duality \cite{PR85}, three-point function QCD Sum Rules 
\cite{threepoint} and the $1/N_c$ ($N_c =$ number of colours) 
expansion \cite{Bkchiral,Bkphysical}. 
Many of these determinations give results for 
$B_K$ only in the chiral limit, since that is a very well established limit 
in the continuum that simplifies the calculation considerably. Chiral 
corrections however are essential to get a value of $B_K$ 
useful for phenomenologists, since they represent more than $50\%$ 
of the final number \cite{Bkphysical}.      
In fact there is only one recent determination of 
the physical $B_K$ \cite{Bkphysical}, although there is 
work in progress to calculate chiral corrections to this parameter in the 
framework of the $1/N_c$ expansion \cite{BGP06,JPR05}.

The fourth technique that has been used to calculate $B_K$ is   
lattice gauge theory. It offers model independent results and has the 
potential to reduce the error to the level required 
by phenomenology. Reviews of lattice determinations of $B_K$ in the 
quenched approximation and 
some preliminary results including quark vacuum polarization effects 
can be found in \cite{reviewslattice}.
  
So far, the value of $B_K$ that the phenomenologists 
have been using in their studies of the UT  
\cite{Charles:2004jd} is a lattice result by the JLQCD collaboration 
\cite{JLQCD97} which uses unimproved staggered quarks in the quenched 
approximation. That is the most complete analysis of $B_K$ with lattice 
techniques to date. The value given in \cite{JLQCD97} is 
$B_K^{\overline{MS}-NDR}(2~\GeV)=0.628\pm0.042$, which 
corresponds to $\hat B_K=0.86\pm0.06$. This number is the result from 
the extrapolation to the continuum limit of 
the values obtained for seven lattice spacings, with both gauge invariant and 
non-invariant operators -see section \ref{operatorsdef}-  
and the error includes an estimate of order $\alpha^2$ and $a^2$ corrections. 
Finite volume effects were also studied for two different values of $a$. 

The main source of uncertainty in the JLQCD calculation (not included 
in the quoted uncertainty) is the unknown error 
from quenching, which could be as large as 15\% according to the Chiral 
Perturbation Theory (CHPT) estimate performed in \cite{Sharpe97}. 
In order to have a prediction at a few percent level it is thus 
necessary to perform unquenched calculations of $B_K$ 
that eliminate the quenching uncertainties. Such calculations have now 
become feasible.

Another feature of the 
calculation in \cite{JLQCD97} is that it is affected by large 
scaling violations. We will see that the scaling behaviour is 
much better using improved staggered actions instead of the 
standard unimproved staggered action used by the JLQCD collaboration.

A third way of improving the JLQCD calculation would be 
to incorporate $SU(3)$  
breaking effects by using kaons made up of non-degenerate quarks, 
instead of degenerate quarks with $m_s/2$. 
However, these effects are not 
expected to be important as first estimated using CHPT by Sharpe 
\cite{Sharpe97} (the effects were estimated to be $\sim 5\%$)  
and as the preliminary results from unquenched domain wall 
\cite{domain2flav} and staggered \cite{BKL05} quarks seem to 
indicate (the authors found differences 
$\sim 3\%$ between the degenerate and non-degenerate results). 

The goal of this work is to perform a calculation of $B_K$ 
including quark vacuum polarization effects, which will  
eliminate the irreducible and unknown quenching uncertainty. 
In this calculation we will use improved staggered fermions that have been 
proved to reduce the large $\order(a^2)$ discretization errors generated by 
the taste-changing interactions. 

Preliminary results from this study were presented in \cite{Gamiz:2004qx}. The 
matching coefficients needed in the calculation of the renormalized 
$B_K$ with the action used in our unquenched simulations were 
not available at that time, so 
an approximate renormalization was performed in order to get those preliminary 
results. The correct renormalization coefficients have since been 
calculated \cite{BGM05} and have been used to 
obtain the results reported in \cite{procBK05} and in the present article  
-see next section for more details about the renormalization process.

Other recent preliminary unquenched results can be found in 
\cite{domain2flav,Flynnetal,Mesciaetal,Cohen05,BKL05,KBL05}.

\section{Theoretical Framework}

\label{sec:theory}

In this section we briefly describe the fermion formalism we are going 
to use in this work, the staggered formalism, as well as 
the fermion and gauge actions used in the quenched and unquenched  
calculations reported here. We also set up the notation for the lattice 
operators needed in our study and justify our choice of external states.

\subsection{Staggered fermions}

When the quark action is written in terms of staggered fermions, 
it becomes spin diagonal and we can drop three of the four components 
of the staggered field $\chi$. The standard (unimproved) form of the 
staggered fermion action is thus 
\be\label{stagaction}
{\cal S}_{f}^{unim.} = \sum_n\left[\frac{1}{2}\sum_\mu \eta_\mu(n)\left(
\bar\chi(n)U_\mu(n)\chi(n+\hat \mu)-\bar\chi(n+\hat\mu)U_\mu^\dagger(n)
\chi(n)\right)+m\bar\chi(n)\chi(n)\right]\,,
\ee
where $n=(n_1,n_2,n_3,n_4)$ parametrizes the lattice site,  
$\eta_\mu(n)=(-1)^{n_1+\dots+n_{\mu-1}}$ and $\chi$ is a 
$3(colours)\times 1(spin)$ component object. The relation between 
the staggered field $\chi$ and the naive fermion field $\Psi$ at each 
lattice site is
\be
\Psi(n)=\Gamma_n\chi(n)\quad\quad\quad
\bar\Psi(n)=\bar\chi(n)\Gamma^\dagger_n\, ,
\ee
with
\be
\Gamma_n = (\gamma_1)^{n_1}(\gamma_2)^{n_2}(\gamma_3)^{n_3}
(\gamma_4)^{n_4}\, .
\ee

The staggered action has a remnant of chiral symmetry which ensures that 
the Goldstone boson pion mass vanishes at $ma=0$. 
But the main advantage of staggered fermions is that they are computationally 
very efficient and unquenched simulations with three flavours of sea  
quarks are possible at present with lighter sea quark masses than those 
achieved with other fermion formulations. 

The fundamental disadvantage suffered by staggered fermions is 
the fact that each flavour field comes in four different tastes. In 
the continuum limit the four tastes are degenerate and extra copies 
can be removed by hand. For sea quarks this involves  
taking the fourth-root of the quark determinant in unquenched simulations 
and, in perturbation theory, the division of each 
fermion loop by a factor of 4. The validity of the fourth root procedure 
has not been rigorously proven but tests of it are encouraging \cite{4roottrick}.
At non-zero lattice spacing the situation 
is more complicated due to the existence of interactions of 
$\order(a^2)$ that violate the taste symmetry and are potentially dangerous. 
However, the negative effects of these taste-changing interactions 
can be reduced by adding improvement terms to the action as   
pointed out in \cite{impactions}. This issue is discussed in more detail in 
the following section.

\subsection{Improved staggered actions}

\label{sec:improv}

Large $\order(a^2)$ discretization errors have been found in the 
calculation of masses and matrix elements using unimproved 
staggered fermions. The origin of these anomalously large 
discretization errors, as well as other effects, such as the large 
size of perturbative corrections, are the taste changing interactions 
that break the taste symmetry at nonzero lattice spacing. 
These (unphysical) interactions 
can be systematically removed using the method of Symanzik and 
replacing the gauge link $U$ in (\ref{stagaction}) by 
a fat link $V$, that is a weighted combination of different staples. 
For studies of the effects of using fat links see 
\cite{DGHK02,fatlinks,OTS99}.

In the quenched simulations in Section \ref{imversusunim}, we use two 
different improved actions, the HYP  and the Asqtad. The construction 
of the HYP \cite{HYP} action involves three levels of APE smearing 
with projection onto SU(3) at each level. The smearing is restricted 
in such a way that each fat link includes contributions only 
from thin links belonging to hypercubes attached to the original 
link.

The Asqtad action we analyze here and use in our unquenched simulations 
has been extensively used in the past for simulating light sea and 
valence quarks \cite{Asqtad}. 
It has been designed to reduce the taste symmetry breaking 
effects and remove all other $\order(a^2)$ discretization errors. 
The virtue of this improved action is that it 
allows for precise calculations with light sea quarks. 

Together with the fat links, the other ingredients that differentiate 
the fermion part of the Asqtad action from the unimproved one are the 
so-called Lepage corrections and the Naik term. 
The Asqtad action, written in terms of four component  
\emph{naive} fermions, has the form
\be
\label{asqtadact}
{\cal S}_f^{Asqtad} = a^4 \sum_x \left\{ \psibar (x) \,
\left[ \sum_\mu \gamma_\mu
{\frac{1}{a}} \left( \nabla^\prime_\mu 
 \,- \, \frac{1}{6} \nabla_\mu^{3-link} \right) 
\; + \; m \right] \Psi(x) \right\} \, ,
\ee
with
\begin{eqnarray}
\nabla^{3-link}_\mu \Psi(x) &=& (\nabla_\mu)^3 \bigg|_{\rm tadpole \; improved}
\Psi(x) \nonumber \\
 &=& \frac{1}{8} \left\{ \frac{1}{u_0^3} \, \left[ 
UUU \,\Psi(x + 3 a_\mu) - U^\dagger U^\dagger U^\dagger \,\Psi(x - 3 a_\mu) 
\right] \right.  \nonumber \\
  & & \quad \left. - \frac{3}{u_0} \left[U \,\Psi(x + a_\mu) - U^\dagger 
\,\Psi(x - a_\mu \right] \right\} \, ,
\end{eqnarray}
and $\nabla_\mu'$ being the usual covariant derivative 
with the thin link variable $U_{\mu}$ replaced by an updated variable 
$V_{\mu}'(x) ~\equiv~ V_\mu(x) - \sum_{\rho\ne\mu} 
\frac{(\nabla^{\ell}_\rho)^2}{4} U_\mu(x)$. In this expression $V_\mu(x)$ 
is the fat link 
\ba
V_\mu(x)\equiv \prod_{\rho\ne \mu}\left(1+\frac{\nabla^{\ell,(2)}_\nu}{4}
\right)\left.\right|_{{\rm symmetrized}}U_\mu(x)
\ea
and the second term is the Lepage term that removes 
a low momentum $\order(a^2)$ error. For the exact definitions of 
the derivatives $\nabla^{\ell}_\nu$ and $\nabla^{\ell,(2)}_\nu$ 
see, for example, \cite{GSW03}. 

The Asqtad fermion action is coupled in the unquenched simulations 
to an SU(3) gluonic action that is one-loop  
Symanzik improved after tadpole improvement \cite{MILC01,impgluon}. 
In the quenched calculations, instead of using the improved gluon action, 
we use the usual unimproved Wilson glue action.

\subsection{Definition of the operators}

\label{operatorsdef}

To construct the four-fermion operators, we collect the staggered fields 
$\chi$ into a set of Dirac fields $q(2N)$ that live on the even 
lattice sites and are spread over a unit hypercube \cite{coordinatespace}
\be
q(2N)_{\alpha i} = \frac{1}{8} \sum_A(\Gamma_A)_{\alpha i}
\chi(2N+A)\,,
\ee
where $\alpha$ and $i$ are the Dirac and taste indices respectively, and 
with A running over the vertices of a hypercube ($A_\mu=0$ or $1$, 
$\mu=1,\dots,4$). The bilinear quark operators with spin structure 
$\gamma_S=\Gamma_S$ and taste structure $\xi_T=\Gamma_T^*$ are defined by 
\be\label{bilindef}
{\cal O_{ST}} = \bar q(2N)(\gamma_S\otimes\xi_T)q(2N)=\frac{1}{16}
\sum_{A,B}\bar\chi(2N+A)\chi(2N+B)\frac{1}{4}
{\rm tr}\left(\Gamma_A^\dagger\gamma_S\Gamma_B\Gamma_T^\dagger\right)
\ee

The four-quark operators can be build with the bilinears in (\ref{bilindef}) 
considering two different contractions of the colour indices
\ba\label{4fermiondef}
{\cal O}_1 &=& \bar q^a(\gamma_{S_1}\otimes\xi_{F_1})q^b\cdot
\bar q^b(\gamma_{S_2}\otimes\xi_{F_2})q^a = \left(\frac{1}{16}\right)^2
\sum_{ABCD}\bar\chi_A^a\frac{1}{4}
{\rm tr}\left(\Gamma_A^\dagger\gamma_{S_1}\Gamma_B\Gamma_{T_1}^\dagger\right)
\chi_B^b\cdot\bar\chi_C^b\frac{1}{4}
{\rm tr}\left(\Gamma_A^\dagger\gamma_{S_1}\Gamma_B\Gamma_{T_1}^\dagger\right)
\chi_D^a\,,
\nonumber\\
{\cal O}_2 &=& \bar q^a(\gamma_{S_1}\otimes\xi_{F_1})q^a\cdot
\bar q^b(\gamma_{S_2}\otimes\xi_{F_2})q^b = \left(\frac{1}{16}\right)^2
\sum_{ABCD}\bar\chi_A^a\frac{1}{4}
{\rm tr}\left(\Gamma_A^\dagger\gamma_{S_1}\Gamma_B\Gamma_{T_1}^\dagger\right)
\chi_B^a\cdot\bar\chi_C^b\frac{1}{4}
{\rm tr}\left(\Gamma_A^\dagger\gamma_{S_1}\Gamma_B\Gamma_{T_1}^\dagger\right)
\chi_D^b\,,
\ea
where we have suppressed the hypercube label $2N$ for simplicity. 
The operators in (\ref{4fermiondef}) are known as one-colour-trace 
and two-colour-trace operators respectively. 

To make these operators gauge invariant we insert gauge link factors 
connecting the quark fields according to 
\ba\label{gaugeinv}
{\cal O}_1 &=& \left(\frac{1}{16}\right)^2
\sum_{ABCD}\bar\chi_A^{a\,n_{f_1}}\frac{1}{4}
{\rm tr}\left(\Gamma_A^\dagger\gamma_{S_1}\Gamma_B\Gamma_{T_1}^\dagger\right)
\chi_B^{b\,n_{f_2}}\cdot\bar\chi_C^{c\,n_{f_3}}\frac{1}{4}
{\rm tr}\left(\Gamma_A^\dagger\gamma_{S_1}\Gamma_B\Gamma_{T_1}^\dagger\right)
\chi_D^{d\,n_{f_4}}\cdot  U_{AD}^{ad}U_{CB}^{cb}\,,
\nonumber\\
{\cal O}_2 &=& \left(\frac{1}{16}\right)^2
\sum_{ABCD}\bar\chi_A^{a\,n_{f_1}}\frac{1}{4}
{\rm tr}\left(\Gamma_A^\dagger\gamma_{S_1}\Gamma_B\Gamma_{T_1}^\dagger\right)
\chi_B^{b\,n_{f_2}}\cdot\bar\chi_C^{c\,n_{f_3}}\frac{1}{4}
{\rm tr}\left(\Gamma_A^\dagger\gamma_{S_1}\Gamma_B\Gamma_{T_1}^\dagger\right)
\chi_D^{d\,n_{f_4}}\cdot U_{AB}^{ab}U_{CD}^{cd}\,.
\ea
Here, we suppress again the hypercube label $2N$ for simplicity and 
add superscripts $n_{f_i}$ to label the different continuum flavours. 
Notice that we consider fat links only in the action and 
those that we introduce in the operators are thin links. 
The only improvement in the operators we consider is tadpole improvement.   
This we carry out by dividing the gluon fields by appropriate 
factors of the mean link $u_0$ defined as the fourth root of the 
average plaquette. The operators used in the calculation 
with improved actions are thus exactly the same as those used in the 
unimproved simulations. In this work we also use  
gauge non-invariant operators, that do not 
incorporate gauge link factors but instead are calculated on 
gluon configurations fixed to Landau gauge. 

In the transcription of the continuum operators involved in the calculation of 
$B_K$ to the staggered ones we have to take into account the taste degree 
of freedom. We choose the external kaon to have taste structure 
$\xi_T=\gamma_5$ (and thus fix the taste structure of the vector and axial 
operators with nonzero matrix elements), since only the mesons with such 
structure become massless in the chiral limit. In addition,  
in order to avoid mixing with operators with different taste structure,  
we follow the two-spin-trace formalism described in references   
\cite{SPGGK87,Sharpe89} and introduce two sets of valence quarks 
$s_1,\, d_1$ and $s_2,\, d_2$ for each of the bilinears constituting  
the four-fermion operator $Q_{\Delta S=2}$. 
Matrix elements are then taken 
between kaons $K^0_1=\overline s_1d_1$ and $\overline{K^0_2}
=\overline d_2 s_2$. A detailed explanation of the two-spin-trace  
formalism for $B_K$ can be found in \cite{SPGGK87,Sharpe89}.

\section{Perturbative renormalization}

\label{sec:perturbative}

An important step needed in the determination of $B_K$ 
(as well as any renormalized quantity) is the calculation of the 
coefficients that match the lattice matrix elements to the continuum ones. 
The operator $Q_{\Delta S=2}$ in (\ref{effHam}) is renormalized 
multiplicatively if one has a regularization with exact chiral symmetry, 
such as the dimensional regularization in the continuum or Ginsparg-Wilson 
fermions on the lattice. For staggered fermions, 
although full chiral symmetry is 
broken at non-zero lattice spacing, there is a remnant $U(1)$ symmetry 
which allows us to have the mixing under control at a given order 
in perturbation theory. In particular, the one-loop calculation involves 
only four operators, as explained below. 

In this work we use a one-loop perturbative matching to relate the 
lattice and continuum operators. A general one-loop matching of the 
bare lattice operators $O_j^{latt.}$ to the renormalized continuum 
operators $O_i^{cont.}$ at an intermediate scale $\mu$ can be expressed as
\be\label{pertmatching}
O_i^{cont.}(\mu) = O_i^{latt.}+\frac{\alpha_s(q^*)}{4\pi}
\sum_j\left(-\gamma^{ij}_1\ln(\mu a)+C_{ij}\right)O_j^{latt.}
+\order(\alpha_s(q^*)^2)\, ,
\ee
with $\gamma_1^{ij}$ the one-loop anomalous dimensions matrix and 
$C_{ij}$ the one-loop matching coefficients, depending on 
the continuum scheme.  

The operator $Q_{\Delta S=2}$ is the product of 
two $V-A$ currents. The QCD corrections to the bare lattice 
four-quark operators affect the vector and axial parts differently; 
as a consequence currents of the form $V+A$ are generated. A minimal set of 
lattice operators that matches to the continuum,  
closes under renormalization and has non-vanishing $K-\overline K$ matrix 
elements is 
\ba
\label{ques}
 V^{(1)}= (\bar s^a d^b)_V (\bar s^b d^a)_V\,, &
\quad \quad V^{(2)} = (\bar s^a d^a)_V (\bar s^b d^b)_V\, ,\nonumber\\
A^{(1)} = (\bar s^a d^b)_A (\bar s^b d^a)_A\,, & 
\quad \quad  A^{(2)} = (\bar s^a d^a)_A (\bar s^b d^b)_A \,,  
\ea
where $V$ and $A$ are the vector and axial currents with taste structure
$\xi_T=\gamma_5$. The colour indices $a,b$ indicate which fields are
connected by gauge link factors, according to (\ref{gaugeinv}). The 
superscripts $1$ and $2$ indicate that they are one-colour-trace 
or two-colour-trace operators. 

Perturbative calculations and non-perturbative simulations 
have to be done with exactly the same action and operators. 
The one-loop 
lattice to continuum matching coefficients for the 
$\Delta S=2$ four-quark operators in (\ref{ques}) have been calculated 
for unimproved staggered fermions \cite{Ishizuka:1993fs}, as well as 
for HYP staggered fermions \cite{Lee:2003sk}; although this last 
calculation only considered gauge invariant operators. 
The matching calculation for the Asqtad action, with both a Wilson 
gauge action and an improved gauge action as described in Section 
(\ref{sec:improv}), was done perturbatively at one-loop in \cite{BGM05} for 
gauge invariant as well as gauge non-invariant operators for a general gauge.

The calculation in \cite{BGM05} was performed with two independent methods. 
First, the authors evaluated the corresponding one-loop diagrams 
in two different ways: by directly calculating the various diagrams 
and by first separating off the part
which can be inferred from the renormalization of the current
operators. Second, the lattice integrals were evaluated both
algebraically and numerically.  For the algebraic evaluation, the authors 
in  \cite{BGM05} expanded the diagrams around the continuum limit.  
This produces a set of lattice tadpole integrals which one then 
reduces to a minimal set of master integrals using computer algebra. 
The agreement between these two rather different methods provides a 
strong check on their results.

The size of the one-loop corrections to the tree level matching for the 
Asqtad action (with and without improved glue) is very similar to what 
is found with unimproved staggered and other improved staggered actions, 
such as the HYP action. In the calculation of $B_K$ thus the 
reason to use an improved action is not the reduction of the matching  
factors (which are not large in the unimproved calculation) but 
the correction of the bad scaling behaviour, as we have already pointed out.
That is not the case with other weak matrix elements relevant in 
the study of $CP$-violating effects, for which the use of improved 
staggered actions greatly reduces the size of the perturbative 
matching coefficients -see \cite{BGM05} for further discussions 
and references.

Two comments are in order with respect to the results in \cite{BGM05}. 
First, the authors found that the matching coefficients obtained 
for the Asqtad action with improved glue are very similar 
to those with an unimproved glue action -see equations 
(\ref{Cquenched}) and (\ref{Cdynam}) below. They conclude that the 
improvement in the glue action is not crucial in order to reduce 
the size of the perturbative coefficients, as already suggested 
in \cite{LS02}. Another conclusion from that work is that we 
do not expect anomalously large $\order(\alpha_s^2)$ corrections 
using the Asqtad action. 

For completeness, we write here the matrices $C_{ij}$ in 
(\ref{pertmatching}),  taken from \cite{BGM05}, 
that must be used to obtain $\langle\overline{K^0}|
Q_{\Delta S=2}(\mu)|K^0\rangle^{\overline{MS}}$ from the Asqtad bare 
matrix elements in the Appendix \ref{apendice}. They are written in the 
basis $(V^{(1)},V^{(2)},A^{(1)},A^{(2)})$. They are 
\ba\label{Cquenched}
C_{{\rm inv.}}^{{\rm unimp.\, glue}} &= \left (
\begin{array}{cccc}
-14.3796 & 1.3606 & -0.0520  & -0.0394  \\
3      & -19.8263 & -0.0389  & 0.0263 \\
-0.0520  & -0.0394  & -15.1796 & 3.8005 \\
-0.0389  & 0.0263 & 3      & -10.6495 \\
\end{array}
\right )\, ,\nonumber\\
C_{{\rm noninv.}}^{{\rm unimp.\, glue}} &= \left (
\begin{array}{cccc}
-12.3228 & 2.6405 & -0.0520  & -0.0394  \\
3      & -13.4411 & -0.0389  & 0.0263 \\
-0.0520  & -0.0394  & -12.4428 & 3.0005 \\
-0.0389  & 0.0263 & 3      & -12.4811 \\
\end{array}
\right )\,,
\ea
for both gauge invariant and gauge non-invariant operators (in 
the Landau gauge) using an unimproved glue action. And
\ba\label{Cdynam}
C_{{\rm inv.}}^{{\rm imp.\, glue}} &= \left (
\begin{array}{cccc}
-15.1091  & 1.7606 & 0.5080 & 0.2406 \\
3       & -19.3177 & 0.4411 & -0.1337 \\
0.5080  & 0.2406 & -15.8691 & 4.0806 \\
0.4411  & -0.1337  & 3      & -10.8972 \\
\end{array}
\right ) ,\nonumber\\
C_{{\rm noninv.}}^{{\rm imp.\, glue}} &= \left (
\begin{array}{cccc}
-12.0486 & 2.9206 & 0.05080  & 0.2406  \\
 3      & -12.2869 & 0.4411  & -0.13370 \\
0.5080  & 0.2406  & -12.1687 & 3.2806 \\
0.4411  & -0.1337 & 3      & -11.3269 \\
\end{array}
\right )\,,
\ea
for gauge invariant operators  and gauge non-invariant operators 
using the improved gluon action.  
These results are obtained after tadpole improvement with the mean 
link $u_0$ defined as the fourth root of the 
average plaquette. The one-loop contribution to this parameter 
is $u_0^{(1)}=\pi/3$ for the unimproved gluon action 
and $u_0^{(1)}=0.7671$ for the improved gluon action, where $u_0^{(1)}$ 
is defined as $u_0=1-\alpha_s u_0^{(1)}+\order(\alpha_s^2)$.

In addition to the matching for the four-fermion matrix element in 
(\ref{BKdef}) we must also account for the renormalization 
of the axial currents in the denominator of (\ref{BKdef}),  
which is non-vanishing with the Asqtad action for the definition the axial 
current of (\ref{ques}). The axial current 
renormalization  is multiplicative and can be written in the form 
$1+\alpha_s/\pi Z_A$, with  $Z_{A,unimp.}^{inv.}=1.206$,  
$Z_{A,imp.}^{inv.}=1.237$ when using unimproved and improved glue 
respectively with gauge invariant operators, and 
$Z_{A,unimp.}^{noninv}=1.435$, 
$Z_{A,imp.}^{noninv}=1.291$ when using unimproved and improved glue with 
gauge non-invariant operators in the Landau gauge.

\section{Scaling behaviour of the improved staggered actions}

\label{imversusunim}

The first issue we analyze is the impact of using improved staggered actions 
in the calculation of $B_K$, in comparison with the unimproved 
staggered action analyzed in \cite{JLQCD97}. This study is carried out 
in the quenched approximation and for two different improved 
actions: the HYP \cite{HYP} and the Asqtad with Wilson glue.

In the next two subsections we describe the results obtained 
for gauge invariant operators only. We postpone  the discussion on 
the differences found between using gauge invariant and gauge 
non-invariant operators until the last subsection \ref{noninvar}.

\subsection{Simulation details}

\begin{table}[t]
\begin{center}
\begin{tabular}{cccccccc}\hline\hline
$\beta$ & Volume & $n_{confs}$ & $a^{-1}(\GeV)$ & $am_s/2$ 
& $La$~(fm) & $am_K$ & $\alpha_s(1/a)$ \\
\hline
\multicolumn{8}{c}{UKQCD $n_f=0$}\\\hline
5.7 & $12^3\times 24$ & 150 & 0.837(6) & 0.086/0.064 &2.9 & 0.417 & 0.31 \\
5.93 & $16^3\times 32$ & 50 & 1.59(3) & 0.039/0.030 &2.0 & 0.220  & 0.21 \\
\hline
\hline
\end{tabular}
\end{center}
\caption{Parameters in the quenched simulations. The values for $m_s/2$ 
are for the HYP and Asqtad staggered actions respectively.
\label{tablequenched}}
\end{table}

We have two ensembles of 150 and 50 configurations 
at $\beta=5.70$ and $\beta=5.93$ respectively,  
generated using the Wilson gluon action for the 
three staggered fermion actions that 
we are going to analyze, unimproved, HYP and Asqtad. 
The values of the parameters used in the simulations are shown in Table 
\ref{tablequenched}. We choose these parameters 
to be the same as those used by the JLQCD collaboration 
in order to make a clear comparison with their results. 
In particular, we match kaon masses at a given 
$\beta$ to those of the JLQCD collaboration \cite{jlqcdlett}, which fixes 
the strange quark mass to the values listed in Table \ref{tablequenched}. 
The lattice spacings, determined from $m_\rho$, are also taken from 
reference \cite{jlqcdlett}, again for consistent comparison with that work. 
For the same reason, we consider kaons made up with two degenerate 
quarks of $m_s/2$ as in  \cite{jlqcdlett}.

\phantom{}
\phantom{}

\begin{figure}[th]
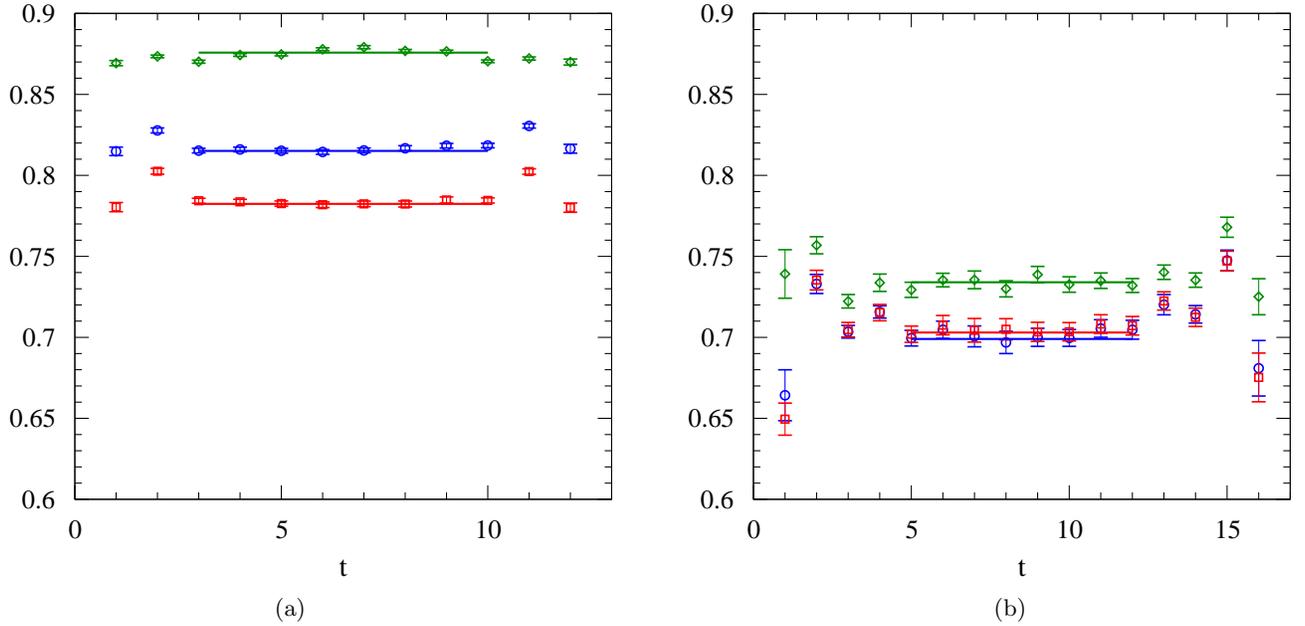

\begin{center}
\begin{tabular}{cc}
\includegraphics[width=0.45\textwidth]{bkbare570} 
\hspace*{0.3cm}&\hspace*{0.3cm}
\includegraphics[width=0.45\textwidth]{bkbare593}
\end{tabular}

\begin{tabular}{cc}
 (a) & \hspace*{9cm}(b)
\end{tabular}
\caption{Values of the bare gauge invariant $B_K$ at $\beta=5.70$ (a) and
$5.93$ (b) for unimproved (green diamonds), HYP (red squares) and 
Asqtad (blue circles) quarks. Solid lines show results from our fits.}
\label{fig:bkbarequenched}
\end{center}
\end{figure}

Within the conventions and parameters we have described above, 
the bare values of $B_K$ 
we obtain are plotted in Figure \ref{fig:bkbarequenched} as a function of 
the timeslice for the three actions and for the two different values of the 
lattice spacing. The $\beta=5.70$ and $\beta=5.93$ results present 
a plateau for $3\le t \le 10$ and $5 \le t \le 12$ respectively, so we make  
fits to a constant over those ranges of values of t. 
The results from those fits are also plotted in Figure 
\ref{fig:bkbarequenched} and the numerical values 
summarized in the left hand side of Table \ref{table:Bkquenched}. 
Individual values for the four four-fermion bare lattice operators 
in (\ref{ques}) normalized to the bare lattice value of the numerator 
in (\ref{BKdef}) are given in Appendix \ref{apendice}.

\begin{table}[th]
\begin{tabular}{ccc}
$B_K^{bare}\, (n_f=0)$& \hspace*{1.5cm} & $B_K^{\overline{MS}-NDR}(2\GeV)\, 
(n_f=0)$\\
\begin{tabular}{ccc}
\hline\hline
\hspace*{0.3cm}$\beta$ \hspace*{0.3cm}& \hspace*{0.3cm}invariant
\hspace*{0.3cm} &\hspace*{0.3cm} non-invariant \hspace*{0.3cm}\\\hline
\multicolumn{3}{c}{$n_f=0$ unimproved}\\\hline
5.7 & 0.876(1)  & 0.895(1)\\
5.93 & 0.734(5) & 0.722(3)\\\hline
\multicolumn{3}{c}{$n_f=0$ Asqtad}\\\hline
5.7 & 0.815(1) & 
0.824(1)\\
5.93 &0.699(7) &
0.708(4) \\\hline
\multicolumn{3}{c}{$n_f=0$ HYP}\\\hline
5.7 &0.782(1)  & - \\
5.93 &0.703(5) & - \\\hline
\end{tabular} & &
\begin{tabular}{ccc}
\hline\hline
\hspace*{0.3cm}$\beta$ \hspace*{0.3cm}& \hspace*{0.3cm}invariant
\hspace*{0.3cm} &\hspace*{0.3cm} non-invariant \hspace*{0.3cm}\\\hline
\multicolumn{3}{c}{$n_f=0$ unimproved}\\\hline
5.7 & 0.816(1)  & 0.841(1)\\
5.93 & 0.720(3) & 0.747(3)\\\hline
\multicolumn{3}{c}{$n_f=0$ Asqtad}\\\hline
5.7 & 0.715(1)  & 0.729(1)\\
5.93 & 0.647(4) & 0.673(4)\\\hline
\multicolumn{3}{c}{$n_f=0$ HYP}\\\hline
5.7 & 0.648(1)  & - \\
5.93 & 0.627(4) & - \\\hline
\end{tabular}
\end{tabular}
\caption{Bare $B_K$ and renormalised $B_K^{\overline{MS}-NDR}(2\GeV)$ for 
both gauge invariant and non-invariant operators in the quenched 
approximation using unimproved, Asqtad and HYP staggered fermions. 
The errors quoted in this table are only statistical.}
\label{table:Bkquenched}
\end{table}

\subsection{Results}

To convert the lattice results to the $\overline{MS}-NDR$ scheme we use 
the one-loop coefficients from \cite{Ishizuka:1993fs}, 
\cite{Lee:2003sk} and \cite{BGM05} as appropriate, with the matching scale 
$\mu$ and the scale for $\alpha_s$ equal to $\mu=q^*=1/a$. 
The values for the coupling constant $\alpha_s(q^*=1/a)$ 
used in the matching process for the different lattice spacings, 
are again taken from \cite{JLQCD97} to provide comparison with that work. 
They use $\alpha_{\overline{MS}-NDR}(1/a)$ as given by 
$\Lambda_{\overline{MS}}=230\MeV$. 
Their values are listed in Table \ref{tablequenched}. After 
this process we obtain the renormalized continuum value 
$B_K^{\overline{MS}-NDR}(1/a)$, that we can run to $2\GeV$ using the two-loop  
running of the continuum renormalization group
\be
B_K (\mu_1) = \frac{C(\mu_2)}{C(\mu_1)}\, B_K(\mu_2)\, , 
\ee
where $C(\mu)$ is the Wilson coefficient given in (\ref{Wilsoncoeff}) 
with $n_f=0$. The renormalization group invariant form $\hat B_K$ can 
be similarly calculated using (\ref{Wilsoncoeff}) and its definition in 
(\ref{hatBk}).

\begin{figure}[t]
\begin{center}
\includegraphics [angle=-90,width=95mm] {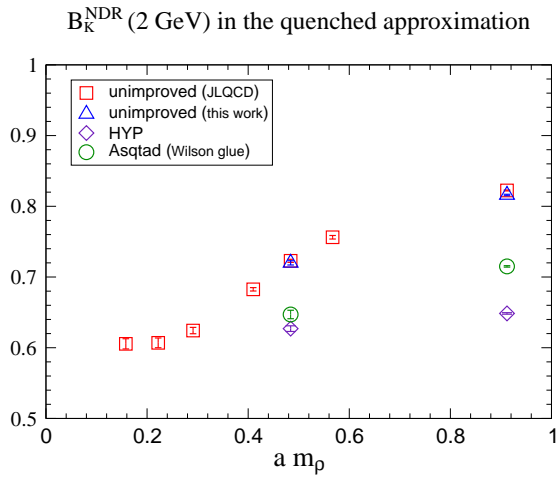}
\end{center}
\caption{Scaling of $B_K^{\overline{MS}-NDR}(2\GeV)$ 
with $a$ for improved staggered actions 
compared to the JLQCD unimproved staggered results. Results in this Figure 
are obtained with gauge invariant operators. \label{quenchedresults}}
\end{figure}

The results we obtain for $B_K^{\overline{MS}-NDR}(2\GeV)$ 
for the different actions and the two lattice spacings are given 
in the right hand side of Table \ref{table:Bkquenched}. 
In the unimproved case the numbers 
for gauge invariant and noninvariant operators  
agree well with those of the JLQCD collaboration \cite{JLQCD97} 
for both lattice spacings.  
The quenched results for gauge invariant operators in Table 
\ref{table:Bkquenched} are plotted  
as a function of the lattice spacing in Figure \ref{quenchedresults}. 
In this figure a clear improvement in the scaling can be seen when 
using improved actions, in particular in the HYP case. We can 
quantify the improvement in the scaling by extrapolating our results 
for $\beta=5.70$ and $\beta=5.93$ to the continuum limit, assuming 
a quadratic dependence on the lattice spacing. In the unimproved case, 
the result from this exercise, 0.682, is incompatible with the 
JLQCD result 0.628, that incorporates an estimate of $\order(a^2)$ 
and $\order(\alpha_s^2)$ corrections. In fact, in the work by the 
JLQCD collaboration \cite{JLQCD97} the authors disregarded the points at 
$\beta=5.70$ and $\beta=5.85$ to perform the extrapolation to the 
continuum limit since these points did not exhibit the quadratic 
dependence in $a$. However, extrapolations of the HYP and Asqtad 
actions for our two lattice spacings give 0.619 and 0.620 for 
$B_K^{\overline{MS}-NDR}(2\GeV)$ in the continuum limit, consistent with 
the JLQCD result in the continuum limit. We expect 
such improved scaling to survive 
unquenching and this therefore allows us to perform reliable unquenched 
calculations with only a few values of the lattice spacing and even 
obtain valuable information from simulations at a single scale. 

The reduction of the discretization errors for staggered fermions using 
improved actions has already been shown for other quantities such as 
hadron masses\cite{fatlinks}.  
In the calculation of $B_K$ it has been 
recently studied in \cite{Bkhyp} with the HYP action, leading to 
the same conclusions as in the present article.

\subsection{Differences between gauge invariant and gauge non-invariant 
operators results}

\label{noninvar}

The final error quoted by the JLQCD collaboration in its 
unimproved staggered study of $B_K$ \cite{JLQCD97} was very much 
enhanced by the differences found between the results obtained 
with gauge invariant and gauge non-invariant operators. 
These differences have their origin in $\order(\alpha_s^2)$ and  
$\order(a^2)$ corrections, which were fitted in \cite{JLQCD97} 
leading to uncertainties of the same size as $3\alpha_s^2$ with 
$\alpha_s=\alpha_{\overline{MS}}(1/a)$. 

We have studied this issue with the unimproved as well as with the 
Asqtad action. A similar analysis with the HYP action is 
not possible since renormalization coefficients for gauge non-invariant 
operators in the Landau gauge are not available.   
The results for the bare $B_K$ as well as the renormalized 
$B_K^{\overline{MS}-NDR}(2\GeV)$ obtained with the gauge non-invariant 
operators defined in Section \ref{operatorsdef} are given in Table 
\ref{table:Bkquenched}. Despite the large improvement in the scaling 
found for the Asqtad action, the differences between gauge invariant 
and noninvariant results are of the same size as those observed in the  
unimproved case. That indicates that these differences are dominated 
not by the $\order(a^2)$ corrections but by the 
$\order(\alpha_s^2)$ corrections, 
as was already pointed out in reference \cite{JLQCD97}. Since the 
improvement of the action does not lead to a reduction of the 
perturbative coefficients in the particular case of the calculation 
of $B_K$, the differences between the results using the two definitions 
of operators are not reduced by using the improved actions. It would 
be necessary to perform a two-loop matching to reduce the uncertainty 
associated with the definition of the operators.

Another conclusion from the comparison of the Asqtad results in Table 
\ref{table:Bkquenched} is that those corresponding to gauge invariant 
operators give values more similar to the JLQCD results as $a\to0$, 
which indicates  that the $\order(\alpha_s^2)$ corrections are 
smaller for these operators. We expect thus more accurate results from 
gauge invariant operators, which is what we use in our unquenched calculation, 
than from the noninvariant ones.

\section{Unquenched value of $B_K$}

\label{sec:DynBK}

We now incorporate quark vacuum polarization effects in 
the calculation of $B_K$ using one of the improved staggered actions 
analyzed in the quenched approximation, Asqtad, 
since the final goal is to eliminate the irreducible systematic 
error associated with quenching that dominates the total uncertainty 
in previous determinations of $B_K$. We use the Asqtad action because there 
are configurations for this action generated with sea masses sufficiently 
small to perform a realistic chiral extrapolation to the physical point 
\cite{LSAB,MILC04}. 
In addition, unquenched simulations using this action have been 
successful in describing a wide range of experimental observables 
with systematic errors of $3\%$ or less \cite{Davies:2003ik}.  

We performed an unquenched calculation of $B_K$ with the Asqtad action 
described in Section \ref{sec:improv}, 
using the configurations from the MILC collaboration with $n_f=2+1$ 
sea flavours \cite{MILC01}. The results reported here correspond 
to the analysis at one lattice spacing with $a=0.125~{\rm fm}$ and two 
different values of the light sea quark masses. The parameters used in the 
unquenched simulations are collected in Table \ref{tabledyn}. 
As in the quenched simulations we use degenerate quark kaons with 
the strange quark mass fixed to its physical value, so no extrapolation 
of the valence quark mass is necessary. 
A difference with the quenched analysis described in the last 
section is that for the unquenched case we only consider gauge invariant 
operators.

\begin{table}
\begin{center}
\begin{tabular}{ccccccc}\hline\hline
$\beta$ & $n_{confs}$  & Volume & $a^{-1}(\GeV)$ 
& $am_{sea}$ & $am_s/2$& $\alpha_V(1/a)$\\
\hline
6.76 &560 & $20^3\times 64$ & 1.605  &0.01/0.05 & 0.02 & 0.47 \\
\hline
6.79 &414 & $20^3\times 64$ & 1.596  &0.02/0.05 & 0.02 & 0.47 \\
\hline
\hline
\end{tabular}
\end{center}
\caption{Parameters in the unquenched simulations. The configurations are 
taken from the MILC collaboration \cite{MILC01}. The lattice spacings 
are taken from \cite{upsilon}\label{tabledyn}. Quarks masses are given in 
the MILC convention, which includes a factor of $u_0$ compared to the standard 
convention.}
\end{table}

\begin{figure}[t]
\begin{center}
\begin{tabular}{cc}
\includegraphics[angle=-90,width=0.45\textwidth]{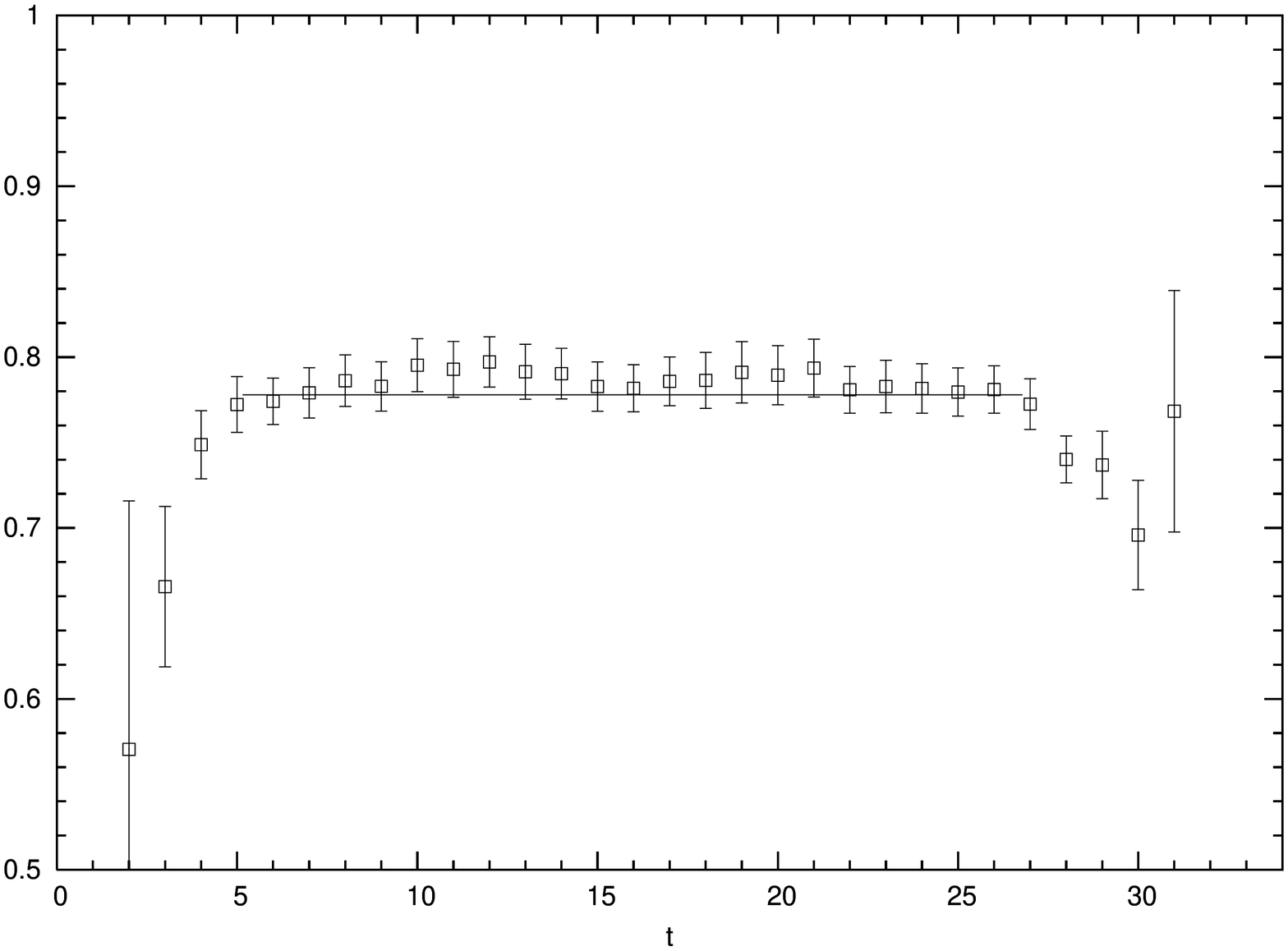} 
\hspace*{0.3cm}&\hspace*{0.3cm}
\includegraphics[angle=-90,width=0.45\textwidth]{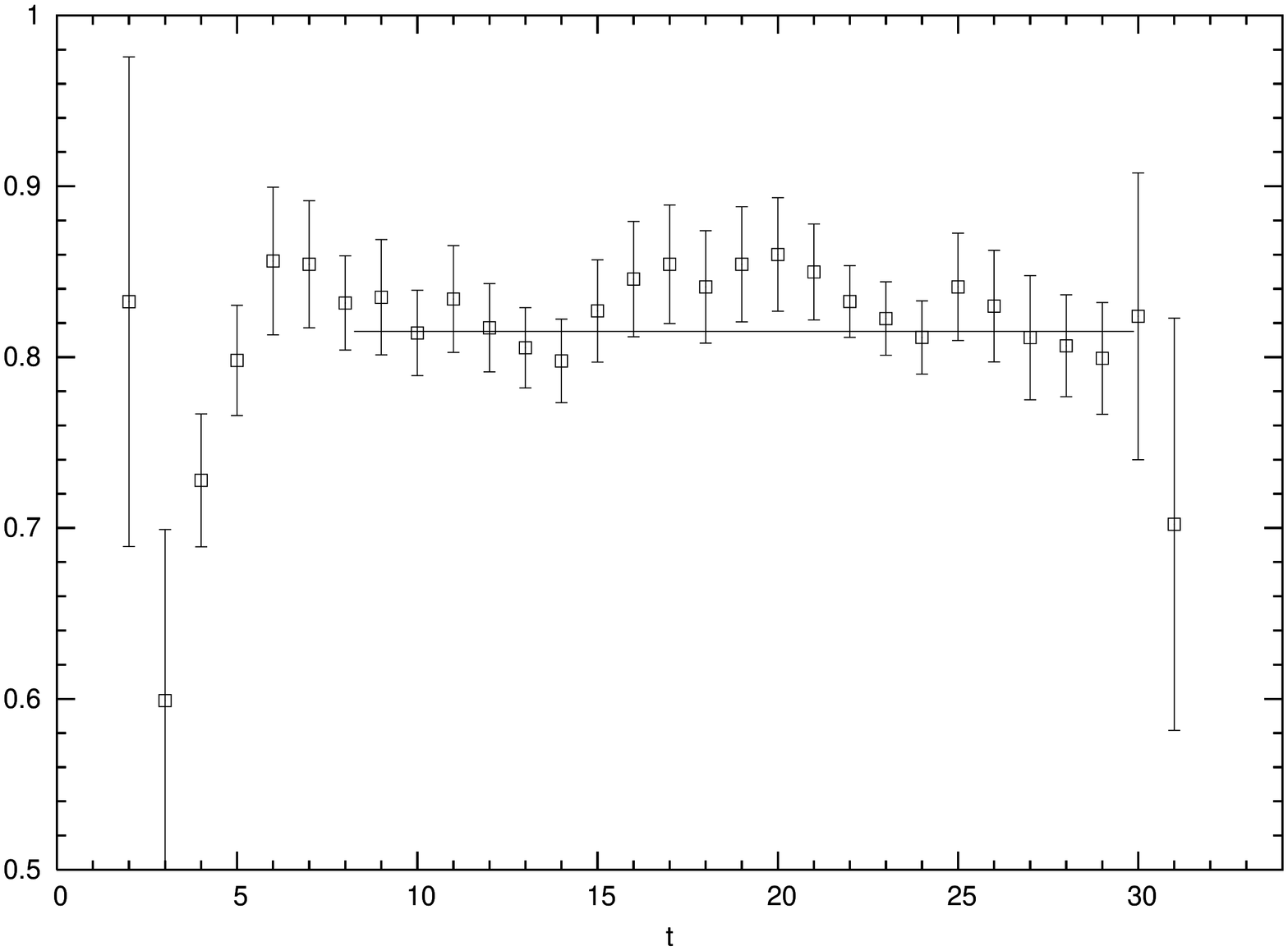}
\end{tabular}

\begin{tabular}{cc}
 (a) & \hspace*{9cm}(b)
\end{tabular}
\caption{Figure showing bare gauge invariant values of $B_K$ as a function 
of the timeslice corresponding to unquenched simulations with 
$m_{sea}=0.01/0.05$ (a) and $m_{sea}=0.02/0.05$ (b).\label{fig:bkbaredyn}}
\end{center}
\end{figure}

A first promising sign in our calculation is that there is little  
contamination from excited states, as can be seen in Figure 
\ref{fig:bkbaredyn}, where we have plotted the bare values of $B_K$ 
as a function of the timeslice. In particular, for $\beta=6.76$, 
for which we have better statistics, we obtain an excellent plateau.   
To get the bare values of $B_K$ given in the Appendix \ref{apendice}, 
we perform fits to a constant over $5\le t \le 27$ for $\beta=6.76$ 
and over $8\le t \le 30$ for $\beta=6.79$. Results from the fits 
over the same ranges for the different four-fermion bare lattice operators 
involved in the calculation, normalized to the bare lattice value 
of the numerator in (\ref{BKdef}), are given in the  Appendix \ref{apendice}. 

The conversion of the values of the bare lattice operators to a value for 
$B_K^{\overline{MS}-NDR}(2\GeV)$ has been done 
perturbatively using the $\order(\alpha_s)$ lattice to continuum 
matching coefficients from \cite{BGM05} collected in the matrix 
(\ref{Cdynam}). In the matching process it is most natural to 
take $\alpha_s$ in the V scheme with values for $N_f=3$ from the recent 
4-loops lattice determination in \cite{Mason:2005zx} 
-see Table \ref{tabledyn}. The scale for $\alpha_s$ is not 
determined here so we consider various reasonable possibilities. One can 
also optimise the scale $\mu$ in equation (\ref{pertmatching}) 
\cite{Nobesetal}, but we have used $\mu=1/a$ throughout this work. 

\begin{table}[t]
\begin{center}
\begin{tabular}{c|cc}\hline\hline
\hspace*{1cm}$m_{sea}$\hspace*{1cm}&\hspace*{1.cm}$B_K^{bare}$ 
\hspace*{0.3cm}& 
\hspace*{0.3cm}$B_K^{\overline{MS}-NDR}(2\GeV)$ \hspace*{0.3cm}\\
\hline\hline
\multicolumn{3}{c}{$n_f=2+1$ Asqtad}\\\hline
0.01/0.05 & 0.785(11) & 0.655(9) \\
0.02/0.05 & 0.815(18) & 0.680(13) \\
\hline
\hline
\end{tabular}
\end{center}
\caption{Values of the bare $B_K$ and the one-loop renormalized 
$B_K^{\overline{MS}-NDR}(2\GeV)$ for the two values of the light sea 
quark masses and the corresponding statistical errors. 
The first number in the row labelled $m_s$ is the mass 
of the degenerate up and down sea quarks and the second one is 
the mass of the strange sea quark. 
\label{table:dynresults}}
\end{table}

\begin{figure}[t]
\begin{center}
\includegraphics [angle=-90,width=85mm] {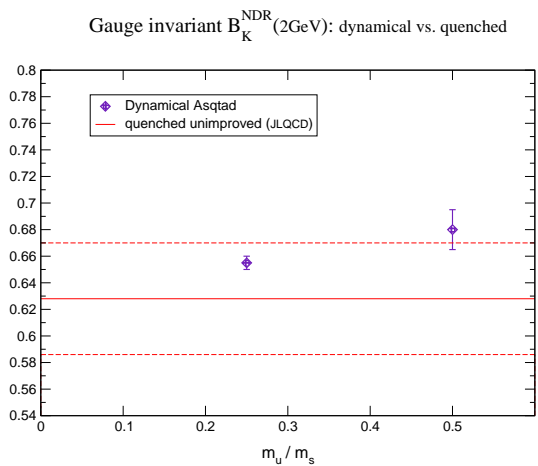}
\end{center}
\caption{Unquenched value of $B_K^{\overline{MS}-NDR}(2\GeV)$ as a function of 
the ratio between 
the light sea quark mass and the (real) strange quark mass. 
The lines represent the quenched results from \cite{JLQCD97}. 
Errors on the points are statistical only. \label{dynresults}}
\end{figure}

The results we obtain for $B_K^{\overline{MS}-NDR}(2\GeV)$  
performing the renormalization with $\alpha_V(1/a)$ 
are shown in Figure \ref{dynresults}. We plot the result as a function 
of the light sea quark mass and  
a decrease of the value of $B_K$ with the reduction of the 
sea quark mass can be appreciated in this figure. For the degenerate 
valence quark case we are analyzing,  
the chiral behaviour of $B_K$ with the sea quark masses is linear, with the 
same coefficient for the strange, up and down sea masses \cite{SW05}.  
We can thus extrapolate our results to the physical $s$ and $u(d)$ masses,   
which yields the result
\be\label{bkvalue}
B_K^{\overline{MS}-NDR}(2\GeV)=0.618(18)(19)(30)(130)\,.
\ee
The first error in (\ref{bkvalue}) is statistical, 
the second is from the extrapolation 
to the physical values of the sea quark masses, 
the third one is from discretizations errors and the final one is from the 
perturbative conversion 
to the $\overline{MS}-NDR$ scheme. The value in (\ref{bkvalue}) 
is equivalent to 
$\hat B_K=0.83\pm0.18$, with $\hat B_K$ defined in (\ref{hatBk}). 
Note that this value of $\hat B_K$ is very 
similar to the previous quenched staggered result in \cite{JLQCD97} 
($\hat B_K=0.86\pm0.06$), so any final conclusion about the enhancement or  
decrease due to the inclusion of quark vacuum polarization 
effects in its calculation would need a significant reduction of the 
error quoted in (\ref{bkvalue}). 

If instead of using $\alpha_s$ in the V scheme to perform the 
renormalization at one-loop, we use $\alpha_S$ in the $\overline{MS}$ 
scheme at a scale $1/a$, the result we find is
\be\label{bkvaluemsbar}
B_K^{\overline{MS}-NDR}(2\GeV)=0.637(19)(20)(31)(82)\,.
\ee
This is the same as the result we obtain when using 
$\alpha_V$ with a scale close to $2/a$.

The total errors in (\ref{bkvalue}) and (\ref{bkvaluemsbar}) 
are dominated by the uncertainty 
associated with possible $\order(\alpha_s^2)$ corrections in the 
lattice to continuum matching process. 
On the finer MILC ensembles on which we are planning to redo our calculation 
as the next step in the improvement of the work 
presented here, the perturbative error bar would be reduced 
since the value of $\alpha_s(1/a)$ would be smaller. Using 
the existing MILC configurations with $a=0.093~fm$,   
the perturbative error could be reduced from $20\%$ to $14\%$  
(or from $14\%$ to $9\%$ if we use $\alpha_{\overline{MS}}(1/a)$ 
in the matching). In view of the improved scaling behaviour 
we obtained in the quenched 
approximation within the Asqtad action described in 
Section \ref{imversusunim}, results for two lattice spacings will be 
enough to perform a reliable continuum extrapolation, reducing 
the discretization errors (third error in (\ref{bkvalue})). 
However, what is really needed 
to reduce the final error in (\ref{bkvalue}) to a few percent level 
is a 2-loop matching or a non-perturbative matching method, that  
eliminates the uncertainty on the possible $\order(\alpha_s^2)$ corrections.

\subsection{Discussion of the perturbative error}

The naive error associated with the perturbative matching that we quote in 
(\ref{bkvalue}) is just the result of multiplying our central value for 
$B_K^{\overline{MS}-NDR}(2\GeV)$ by $\left(\alpha_V(1/a)\right)^2$. 
In the same way, the perturbative error in (\ref{bkvaluemsbar}) is the 
product of $B_K^{\overline{MS}-NDR}(2\GeV)$ in that equation and 
$\left(\alpha_{\overline{MS}}(1/a)\right)^2$. 
At the lattice spacing we are working, the unquenched values 
of $\alpha_V$ are large -see Table \ref{tabledyn}- and translate 
into a $\sim 20\%$ error in the result for $B_K^{\overline{MS}-NDR}(2\GeV)$. 

In principle, we do not expect two-loop corrections as large as $20\%$, since 
for the Asqtad action there are perturbative corrections for both 
numerator and denominator in (\ref{BKdef}) that tend to cancel at any 
order in the expansion. In fact, the shift in the value of $B_K$ 
due to the one-loop renormalization is only $18\%$.    
Further evidence in favour of smaller $\order(\alpha_s^2)$ corrections 
is the fact that the difference between the results we obtain doing 
the matching with $\alpha_s$ at $q^*=1/a$ and $q^*=2/a$, that can 
be taken as an estimate of the uncertainty from the truncation of 
the perturbative series, is around $4\%$. The perturbative uncertainty 
estimated by squaring $\alpha_{\overline{MS}}(1/a)$, $13\%$, then seems more 
realistic. On the other hand, we believe that taking the two-loop error 
to be just the square of the one-loop shift 
\cite{Nierste05} could be an underestimate of the error in the 
absence of other information.

\section{Summary and Conclusions}

Most of the previous lattice calculations of $B_K$, in particular those 
used in the UT analysis, were performed in the quenched approximation. 
This induces a large, 
essentially unknown and irreducible systematic error into 
the result. Precise simulations with sea quarks are necessary in order 
to be able to make full use of the experimental data on $\varepsilon_K$ to 
constrain the CKM matrix. These unquenched  
simulations are feasible with present 
computers using staggered fermions at light sea quark masses. However, 
the unimproved staggered action suffers from large taste-changing interactions 
that generate important scaling corrections, as those found by the JLQCD 
results. We have shown in Section \ref{imversusunim} that the scaling 
behaviour is much better when using improved staggered actions. 

This reduction of the discretization errors, together with the 
existence of unquenched configurations 
with relatively small sea quark masses, makes improved staggered actions an 
ideal choice for accurate calculations of $B_K$ that incorporate 
light quark vacuum polarization effects. 
As a first step in this study we have calculated $B_K$ with the Asqtad 
action in two ensembles at $a=0.125~{\rm fm}$ and with two different 
light sea quark masses. In doing that, we have used the recent results 
for the one-loop matching coefficients in \cite{BGM05}. 
We obtain, using $\alpha_V(1/a)$,
\be\label{BKvalue}
B_K^{\overline{MS}-NDR}(2\GeV)=0.618\pm0.136 \quad\quad{\rm or,\,
 equivalently,\,} \quad\quad\hat B_K=0.83\pm0.18\,,
\ee
or, performing the perturbative matching with $\alpha_{\overline{MS}}(1/a)$, 
\be\label{BKvaluemsbar}
B_K^{\overline{MS}-NDR}(2\GeV)=0.637\pm0.092 \quad\quad{\rm or,\,
 equivalently,\,} \quad\quad\hat B_K=0.85\pm0.12\,.
\ee
Both results are compatible within errors 
with other preliminary unquenched determinations 
\cite{domain2flav,Flynnetal,Mesciaetal,Cohen05,BKL05}.

The error needs to be reduced further. Some reduction can be achieved by 
working on finer lattices and incorporating staggered chiral perturbation 
theory results \cite{SW05}, but a large reduction will require matchings 
coefficients calculated beyond $\order(\alpha_s)$.

Another issues we would like to investigate in the future are 
the impact of $SU(3)$ breaking effects and the chiral limit value 
of this quantity that could be compared to recent continuum 
calculations \cite{Bkchiral}.

\section{Acknowledgments}
This work was supported by European Commission (EC) 
Hadron Physics I3 Contract RII3-CT-2004-506078 and Marie Curie Grant No. 
MEIF-CT-2003-501309, by PPARC, the US Department of Energy 
and the National Science Foundation. Calculations were done at NERSC. 
We thank the MILC collaboration for making their unquenched gauge 
configurations available. We thank UKQCD for their quenched configurations. 
We thank T. Becher and K. Melnikov for useful discussions about 
the renormalization of $B_K$.

\appendix  

\section{Details of calculation results}

\label{apendice}

In this Appendix we collect the different four-fermion bare lattice matrix  
elements involved in the calculation of $B_K$ for all the cases we 
analyze in this work. 

\begin{table}[h]
\begin{tabular}{c|cccccccc}
$\beta$ & $\sum A_i^{(1)}$ & $A_4^{(1)}$ & $\sum A_i^{(2)}$ & $A_4^{(2)}$ 
& $\sum V_i^{(1)}$ & $V_4^{(1)}$ & $\sum V_i^{(2)}$ & $V_4^{(2)}$ \\
\hline\hline
\multicolumn{9}{c}{{\bf Unimproved} $n_f=0$ (gauge invariant operators)}\\
\hline
5.70 &0.026(1)  &0.226(1) &0.012(1) &0.775(1) &-0.143(1) &-0.010(1) &-0.009(1) &
-0.002(1) \\
5.93 & 0.217(5) & 0.332(2) &0.091(2) &0.809(2) & -0.551(7) & -0.079(2) & 
-0.065(1) & -0.0182(4)\\
\multicolumn{9}{c}{\phantom{}}\\
\multicolumn{9}{c}{{\bf Unimproved} $n_f=0$ (gauge non-invariant operators)}\\
\hline
5.70 &0.035(1)  &0.276(1)  &0.014(1)  & 0.772(1)  & -0.176(1) &-0.013(1)  &
-0.010(1)  &-0.002(1) \\
5.93 &0.211(3)  &0.346(1)  &0.080(1)  &0.800(2)  & -0.571(5) & -0.077(1) &
-0.056(1)  &-0.015(1)\\
\end{tabular}
\caption{Values of the gauge invariant and gauge non-invariant bare 
matrix elements in the quenched approximation, shown as a ratio 
with the denominator in (\ref{BKdef}) for unimproved staggered fermions.
\label{tablaunimp}}
\end{table}

\begin{table}[h]
\begin{tabular}{c|cccccccc}
$\beta$ & $\sum A_i^{(1)}$ & $A_4^{(1)}$ & $\sum A_i^{(2)}$ & $A_4^{(2)}$ 
& $\sum V_i^{(1)}$ & $V_4^{(1)}$ & $\sum V_i^{(2)}$ & $V_4^{(2)}$ \\
\hline\hline
\multicolumn{9}{c}{{\bf HYP} $n_f=0$ (gauge invariant operators)}\\\hline
5.70 & 0.048(1)  &0.262(1) &0.016(1) &0.772(1) &-0.285(2) &-0.0194(4) 
&-0.0079(3) & -0.0016(1) \\
5.93 &0.181(8) &0.333(4) &0.065(6) &0.798(5) &-0.540(13) &-0.061(3) &
-0.042(3) &-0.009(1) \\
\multicolumn{9}{c}{\phantom{}}\\
\multicolumn{9}{c}{{\bf HYP} $n_f=0$ (gauge non-invariant operators)}\\
\hline
5.70 & 0.041(1)  & 0.279(1)  &0.014(1)  &0.769(1)   & -0.278(1)  &-0.017(0)   &
-0.008(0)   & -0.001(0)\\
5.93 & 0.164(8)  & 0.344(4)  & 0.053(5) & 0.796(5)  & -0.550(1)  &-0.058(2)&
-0.039(3)   & -0.008(7)\\
\end{tabular}
\caption{Values of the gauge invariant and gauge non-invariant bare 
matrix elements in the quenched approximation, shown as a ratio 
with the denominator in (\ref{BKdef}) for HYP staggered fermions.
\label{tablaHYP}}
\end{table}

\begin{table}[h]
\begin{tabular}{c|cccccccc}
$\beta$ & $\sum A_i^{(1)}$ & $A_4^{(1)}$ & $\sum A_i^{(2)}$ & $A_4^{(2)}$ 
& $\sum V_i^{(1)}$ & $V_4^{(1)}$ & $\sum V_i^{(2)}$ & $V_4^{(2)}$ \\
\hline\hline
\multicolumn{9}{c}{{\bf Asqtad} gauge invariant operators}\\
\hline
\multicolumn{9}{c}{$n_f=0$}\\\hline
5.7 &0.049(6)  &0.285(1) &0.018(1) &0.781(1) &-0.278(1) &
-0.020(1) &-0.016(1) &-0.004(1) \\
5.93 &0.198(6) &0.351(3) &0.077(4) &0.808(4) &-0.596(11) &-0.074(2) &
-0.053(2) &-0.012(1) \\
\hline
\multicolumn{9}{c}{$n_f=2+1$ $m_{sea}=0.01/0.05$ $m_{val}=0.02$}\\\hline
6.76 &0.248(6) & 0.316(5) & 0.076 (2) & 0.885(16) & -0.605(11) & -0.093(2) & 
-0.035(1) & -0.0092(4) \\\hline
\multicolumn{9}{c}{$n_f=2+1$ $m_{sea}=0.02/0.05$ $m_{val}=0.02$}\\\hline
6.79 & 0.244(8) & 0.323(8) & 0.080(3) & 0.940(26) & -0.630(17) & -0.096(2) 
& -0.036(1) & -0.0099(4) \\
\multicolumn{9}{c}{\phantom{}}\\
\multicolumn{9}{c}{{\bf Asqtad} gauge non-invariant operators}\\
\hline
\multicolumn{9}{c}{$n_f=0$}\\\hline
5.7 & 0.047(1)  & 0.284(0)  & 0.017(0)  & 0.777(1)  &-0.267(1)   & -0.018(0)  &
-0.011(0)   & -0.002(0)\\
5.93 &0.164(4) & 0.338(2) & 0.066(2) & 0.797(3) & -0.548(7)&
-0.061(2) & -0.041(2) & -0.009(4)\\
\end{tabular}
\caption{Values of the gauge invariant and gauge non-invariant 
bare matrix elements in the quenched approximation and with  
$n_f=2+1$ sea flavours, shown as 
a ratio with the denominator in (\ref{BKdef}) for Asqtad staggered fermions. 
\label{tablaAsqtad}}
\end{table}

\end{document}